\documentclass[a4paper, conference]{IEEEtran}
\IEEEoverridecommandlockouts 
\ifCLASSINFOpdf
\usepackage[pdftex]{graphicx}
\graphicspath{{./figure/ }}
\DeclareGraphicsExtensions{.pdf,.jpeg,.png}
\else
\usepackage[dvips]{graphicx}
\DeclareGraphicsExtensions{.eps}
\fi

\usepackage{tikz}
\usepackage{pgfplots}
\pgfplotsset{compat=1.15}
\usetikzlibrary{calc,shapes.geometric,shapes.misc,arrows,matrix}
\usepackage[absolute,overlay]{textpos}

\usepackage[font=footnotesize]{caption}
\usepackage[font=footnotesize]{subcaption}
\usepackage{float}
\graphicspath{{figure/}}

\usepackage[utf8]{inputenc}
\usepackage[nolist]{acronym}
\usepackage[cmex10]{amsmath}
\usepackage{mdwmath}
\usepackage{mdwtab}
\usepackage{afterpage}
\usepackage{multirow}
\usepackage{amsthm}
\usepackage{amssymb}
\usepackage{mathtools}
\newtheorem{theorem}{Theorem}

\newtheorem{corollary}{Corollary}

\usepackage{siunitx}
\DeclareSIUnit{\dBi}{dBi}
\DeclareSIUnit{\dBm}{dBm}
\DeclareSIUnit{\dBW}{dBW}
\usepackage{color}
\usepackage{dashrule}
\usepackage{algorithm} 
\usepackage{algpseudocode}
\usepackage{cleveref}
\usepackage{cite}
\floatstyle{plaintop}
\restylefloat{table}
\usepackage{balance}
\usepackage{bm}

\usepackage{amssymb}

\newcommand{\vek}[1]{\ensuremath{\mathbf{#1}}}          

\newcommand{\Exvl}[1]{\ensuremath{\mathbb{E}\left\{#1\right\}}}
\newcommand{\E}{\ensuremath{\mathbb{E}}}

\newcommand{\nvar}{\ensuremath{\sigma_\mathsf{n}^2}}

\newcommand{\NS}{\ensuremath{N_{\mathrm{S}}}}          

\newcommand{\Nt}{\ensuremath{N_{\mathrm{T}}}}

\newcommand{\GTx}{\ensuremath{\zeta_{\mathrm{Tx,dB}}}}
\newcommand{\GRx}{\ensuremath{\zeta_{\mathrm{Rx,dB}}}}

\newcommand{\DA}{\ensuremath{D_{\mathrm{A}}}}
\newcommand{\DS}{\ensuremath{D_{\mathrm{S}}}}

\newcommand{\Ptx}{\ensuremath{P_{\mathrm{Tx}}}}

\newcommand{\aod}{\ensuremath{\theta}}
\newcommand{\aodl}{\ensuremath{\theta_\ell}}

\newcommand{\el}{\ensuremath{^{\mathrm{el}}}}
\newcommand{\az}{\ensuremath{^{\mathrm{az}}}}

\newcommand{\erraoa}{\ensuremath{\upsilon}}

\newcommand{\erraod}{\ensuremath{\xi}}

\newcommand{\steeraod}{\ensuremath{\vek{a}}}
\newcommand{\steeraodest}{\ensuremath{\hat{\steeraod}}}
\newcommand{\Steeraod}{\ensuremath{\vek{A}}}

\newcommand{\xdim}{\ensuremath{\mathsf{x}}} 
\newcommand{\ydim}{\ensuremath{\mathsf{y}}} 
\newcommand{\zdim}{\ensuremath{\mathsf{z}}}

\newcommand{\cfaod}{\ensuremath{\varphi_{\mathsf{\xi}}}}
\newcommand{\cfgau}{\ensuremath{\varphi_{\mathsf{G}}}}
\newcommand{\cfuni}{\ensuremath{\varphi_{\mathsf{U}}}}

\newcommand{\corrmtx}{\ensuremath{\vek{R}}}
\newcommand{\eigvec}{\ensuremath{\boldsymbol{\psi}}}

\newcommand{\corrmtxl}{\ensuremath{\corrmtx_{\steeraod_\ell}}}
\newcommand{\corrmtxlx}{\ensuremath{\corrmtx_{\steeraod_\ell^\xdim}}}
\newcommand{\corrmtxly}{\ensuremath{\corrmtx_{\steeraod_\ell^\ydim}}}
\newcommand{\corrmtxi}{\ensuremath{\corrmtx_{\steeraod_i}}}

\newcommand{\C}{\ensuremath{\mathbb{C}}}

\newcommand{\fc}{\ensuremath{f_{\text{c}}}}

\newcommand{\vekx}{\ensuremath{\vek{x}}}
\newcommand{\veks}{\ensuremath{\vek{s}}}

\newcommand{\vekv}{\ensuremath{\vek{v}}}
\newcommand{\veky}{\ensuremath{\vek{y}}}

\newcommand{\vekn}{\ensuremath{\vek{n}}}

\newcommand{\PC}{\ensuremath{\vek{G}}}
\newcommand{\pc}{\ensuremath{\vek{g}}}

\newcommand{\hugu}{\ensuremath{\vek{h}_{\ell}^H\vek{g}_{\ell}}}
\newcommand{\hvgu}{\ensuremath{\vek{h}_{i}^H\vek{g}_{\ell}}}
\newcommand{\hugv}{\ensuremath{\vek{h}_{\ell}^H\vek{g}_{i}}}

\newcommand{\vekH}{\ensuremath{\vek{H}}}

\newcommand{\vekh}{\ensuremath{\vek{h}}}

\newcommand{\vekI}{\ensuremath{\vek{I}}}

\newcommand{\vekU}{\ensuremath{\vek{U}}}
\newcommand{\vekV}{\ensuremath{\vek{V}}}

\newcommand{\sinr}{\ensuremath{\Gamma}}
\newcommand{\slnr}{\ensuremath{\bar{\gamma}}}

\newcommand{\diag}[1]{\ensuremath{\operatorname{diag}\left(#1\right)}}

\newcommand{\tr}[1]{\ensuremath{\operatorname{tr}\left\{#1\right\}}} 

\newcommand{\sinc}[1]{\ensuremath{\operatorname{sinc}\left(#1\right)}}

\newcommand{\nullvec}{\boldsymbol{0}}   

\begin{acronym}
    \acro{uav}[UAV]{unmanned aerial vehicle}
	\acro{tx}[VSAT]{very small aperture terminal}
	\acro{rx}[GS]{ground station}
	\acro{dl}[DL]{downlink}
	\acro{ul}[UL]{uplink}
	\acro{mimo}[MIMO]{multiple-input-multiple-output}
	\acro{ntn}[NTN]{non-terrestrial network}
	\acro{geo}[GEO]{geostationary orbit}
	\acro{meo}[MEO]{medium Earth orbit}
	\acro{leo}[LEO]{low Earth orbit}
	\acro{haps}[HAPS]{high altitude platform station}
	\acro{qos}[QoS]{quality of service}
	\acro{dip}[DiP]{distributed precoding}
	\acro{pspc}[PAPPC]{per-acces point power constraint}
	\acro{kkt}[KKT]{Karush-Kuhn-Tucker}
	\acro{svd}[SVD]{singular value decomposition}
	\acro{mse}[MSE]{mean-squared error}
	\acro{mmse}[MMSE]{minimum mean-squared error}
	\acro{mvdr}[MVDR]{minimum variance distortionless response}
	\acro{zf}[ZF]{zero-forcing}
	\acro{mrt}[MRT]{maximum ratio transmission}
	\acro{mrc}[MRC]{maximum ratio combining}
	\acro{sic}[SIC]{successive interference cancellation}
	\acro{wrt}[w.\,r.\,t.]{with respect to}
	\acro{iid}[i.\,i.\,d.]{independent and identically distributed}
	\acro{snr}[SNR]{signal-to-noise ratio}
	\acro{sinr}[SINR]{signal-to-interference-and-noise ratio}
	\acro{slnr}[SLNR]{signal-to-leakage-and-noise ratio}
	\acro{csi}[CSI]{channel state information}
	\acro{csit}[CSIT]{\ac{csi} at the transmitter}
	\acro{csir}[CSIR]{\ac{csi} at the receiver}
	\acro{los}[LOS]{line-of-sight}
	\acro{nlos}[NLOS]{non-\ac{los}}
	\acro{pl}[PL]{path loss}
	\acro{fspl}[FSPL]{free space \ac{pl}}
	\acro{aod}[AoD]{angle of departure}
	\acrodefplural{aod}{angles of departure}
	\acro{aoa}[AoA]{angle of arrival}
	\acrodefplural{aoa}{angles of arrival}
	\acro{ula}[ULA]{uniform linear array}
	\acro{ura}[URA]{uniform rectangular array}
	\acro{ecef}[ECEF]{Earth-centered, Earth-fixed }
	\acro{dft}[DFT]{discrete Fourier transform}
	\acro{sota}[SotA]{state-of-the-art}
	\acro{pdf}[PDF]{probability density function}
	\acro{cf}[CF]{characteristic function}
	\acro{fdd}[FDD]{frequency division duplex}
	\acro{isl}[ISL]{inter-satellite link}
\end{acronym}

\allowdisplaybreaks

\usepackage{moreverb}
\immediate\write18{texcount -inc -incbib 
-sum main.tex > /tmp/wordcount.tex}

\begin{document}
\title{Robust Precoding via Characteristic Functions for VSAT to Multi-Satellite Uplink Transmission}

\author{
	\IEEEauthorblockN{
		Maik Röper\IEEEauthorrefmark{1}, Bho Matthiesen\textsuperscript{\IEEEauthorrefmark{1}\IEEEauthorrefmark{2}}, Dirk Wübben\IEEEauthorrefmark{1}, Petar Popovski\textsuperscript{\IEEEauthorrefmark{3},\IEEEauthorrefmark{2}}, and Armin Dekorsy\IEEEauthorrefmark{1}}%
	\IEEEauthorblockA{\IEEEauthorrefmark{1} Gauss-Olbers Center, c/o University of Bremen, Dept. of Communications Engineering, 28359 Bremen, Germany\\
		\IEEEauthorrefmark{2} University of Bremen, U Bremen Excellence Chair, Dept.\ of Communications Engineering, 28359 Bremen, Germany\\
		\IEEEauthorrefmark{3} Aalborg University, Department of Electronic Systems, 9220 Aalborg, Denmark\\
		Email: \{roeper, matthiesen, wuebben, dekorsy\}@ant.uni-bremen.de, petarp@es.aau.dk}
}

\maketitle

\begin{abstract}
	The uplink from a very small aperture terminal (VSAT) towards multiple satellites is considered, in this paper. VSATs can be equipped with multiple antennas, allowing parallel transmission to multiple satellites. A low-complexity precoder based on imperfect positional information of the satellites is presented.
	The probability distribution of the position uncertainty and the statistics of the channel elements are related by the characteristic function of the position uncertainty.
	This knowledge is included in the precoder design to maximize the mean signal-to-leakage-and-noise ratio (SLNR) at the satellites.
	Furthermore, the performance w.r.t. the inter-satellite distance is numerically evaluated. It is shown that the proposed approach achieves the capacity for perfect position knowledge and sufficiently large inter-satellite distances.
	In case of imperfect position knowledge, the performance degradation of the robust precoder is relatively small.
\end{abstract}

\section{Introduction}
In future mobile networks, the terrestrial 
infrastructure will be extended by \acp{ntn}, consisting of \acp{uav}, \acp{haps} and communication satellites, forming a global heterogeneous 3D network.
An integral component to these spatial networks are constellations of small satellites in \acp{leo} \cite{Kodheli.etal.2021,3GPP.TR.38.863}. This is because, in comparison to satellites in \ac{meo} and \ac{geo}, these constellations have reduced deployment costs and acceptable transmission delays \cite{Di.Song.Li.Poor.2019,Leyva-Mayorga2020}.

Recent results examine the application of \ac{mimo} technologies to simultaneously communicate with multiple satellites \cite{Arapoglou.etal.2011,Goto.Shibayama.Yamashita.Yamazato.2018,Hofmann.Storek.Schwarz.Knopp.2016,Richter.Bergel.Noam.Yair.2020}. In combination with formation flying techniques, which allow multiple satellites to move in clusters and act as a single entity \cite{Verhoeven.Bentum.Monna.Rotteveel.Guo.2011,Radhakrishnan.etal.2016,Liu.Zhang.2018}, swarms of small satellites can form huge virtual antenna arrays with the promise of massive spectral efficiency improvements.

In \cite{Roeper.wcnc.2022,Roeper.twc.2022}, optimal geometrical conditions for satellite swarms \ac{wrt} the channel capactiy are derived.
Additionally, a low-complexity linear transceiver design for the downlink from a satellite swarm towards a \ac{mimo} ground terminal has been proposed for the considered scenario.
It has been observed that the derived transceiver achieves the channel capacity for sufficiently large inter-satellite distances and perfect position knowledge.

In case of imperfect position knowledge, the statistics of the uncertainty can be included to design a robust precoder.
In \cite{Chu.Chen.Zhong.Zhang.2021} robust precoding under imperfect phase knowledge for multi-user downlink scenario is presented. The precoder is designed to achieve a given \ac{sinr} per user with minimum transmit power. In order to deal with the phase uncertainty, the \ac{sinr} constraint is relaxed to be satisfied either only in mean, or with a given probabilistic. 
Similarly, in \cite{Wang.Gao.Ding.Lei.You.Chan.Gao.2021,Liu.Yang.Li.Li.Feng.2022} robust precoder with a relaxed average \ac{sinr} constraint in the presence of phase uncertainties are presented. The precoder in \cite{Wang.Gao.Ding.Lei.You.Chan.Gao.2021} allows a trade-off between spectral and energy efficiency, while in \cite{Liu.Yang.Li.Li.Feng.2022} the energy efficiency for a hybrid precoder is maximized.
A precoder under the strict constraint to ensure an \ac{sinr} above a given threshold in case of bounded position uncertainty is given in \cite{Lin.Lin.Huang.Cola.Zhu.2019}. There, different objective functions are numerically optimized, while the uncertainty region is sampled to obtain a finite set of constraints.
In \cite{Schwarz.2018} another robust precoder for the multi-user downlink is presented to maximize the signal power of a user while keeping the interference leakage to other users within a bounded uncertainty region below a given threshold.
A different approach to deal with position uncertainties is been presented in \cite{Roeper.twc.2022}. There, it is shown that the \ac{cf} of the position error is related to the autocorrelation matrix of the channel.

In this paper, we utilize the \ac{cf} of the position uncertainty to present a novel robust precoder to maximize the mean \ac{slnr} \cite{Sadek.Tarighat.Sayed.2007} for \ac{tx} to multi-satellite uplink applications.
Therefore, different to \cite{Lin.Lin.Huang.Cola.Zhu.2019,Schwarz.2018}, our precoder is suitable for arbitrary probability distributions of the receiver positions. Furthermore, we obtain an analytic solution, and therefore the computational complexity is relatively low, as the precoder only requires an eigendecomposition. 
Furthermore, we show numerically that the proposed precoder achieves the capacity for sufficiently large inter-satellite distances and perfect position knowledge of the satellites at the \ac{tx}.

The rest of the paper is organized as follows. Next, in Section~\ref{sec:system}, the general system model and the performance benchmark for perfect \ac{csi} is presented. In Section~\ref{sec:pc} the proposed robust precoder is presented and  numerically evaluated in Section~\ref{sec:simulation}. Finally, Section~\ref{sec:conclusion} concludes the paper.
%
%

\section{System Model and Performance Bounds}\label{sec:system}
\begin{figure}
	\centering
	\tikzstyle{block} = [draw, fill=white, rectangle, 
    minimum height=3em, minimum width=3.3em]

\newcommand{\Nsat}{N_{\mathrm{S}}}

\tikzset{array/.pic ={
	\draw[thick, fill, opacity=0.4] (0,0) -- (-0.3,-0.5) -- (0.3,-0.5) -- (0.6,-0) -- (0,0);
	
%
%
%
}}

\tikzset{sat/.pic ={
	\draw[fill, opacity=1] (-0.2,-0.3) -- (-0.2,0.3) -- (0.2,0.3) -- (0.2,-0.3) -- (-0.2,-0.3);
	\draw (0.2,0) -- (0.4,0);
	\draw (-0.2,0) -- (-0.4,0);
	\draw[fill, opacity=0.7] (0.5,-0.2) -- (0.3,0.2) -- (1.1,0.2) -- (1.3,-0.2) -- (0.5,-0.2);
	\draw[fill, opacity=0.7] (-0.3,-0.2) -- (-0.5,0.2) -- (-1.3,0.2) -- (-1.1,-0.2) -- (-0.5,-0.2);
	\draw (0.0,-0.3) -- (0.0,-0.35);
	\pic[rotate=5] () at (-0.15,-0.1) {array};

}}

\tikzset{gs/.pic ={
\draw[thick] (0,0) -- (-0.4,-0.6) -- (0.4,-0.6) -- (0.8,-0) -- (0,0);
	
\draw[fill] (0,0) circle[radius=0.05];
	\draw[fill] (0.2,0) circle[radius=0.05];
	\draw[fill] (0.4,0) circle[radius=0.05];
	\draw[fill] (0.6,0) circle[radius=0.05];
	\draw[fill] (0.8,0) circle[radius=0.05];

	\draw[fill] (-0.13,-0.2) circle[radius=0.05];
	\draw[fill] (0.07,-0.2) circle[radius=0.05];
	\draw[fill] (0.27,-0.2) circle[radius=0.05];
	\draw[fill] (0.47,-0.2) circle[radius=0.05];
	\draw[fill] (0.67,-0.2) circle[radius=0.05];

	\draw[fill] (-0.26,-0.4) circle[radius=0.05];
	\draw[fill] (-0.06,-0.4) circle[radius=0.05];
	\draw[fill] (0.14,-0.4) circle[radius=0.05];
	\draw[fill] (0.34,-0.4) circle[radius=0.05];
	\draw[fill] (0.54,-0.4) circle[radius=0.05];

	\draw[fill] (-0.4,-0.6) circle[radius=0.05];
	\draw[fill] (-0.2,-0.6) circle[radius=0.05];
	\draw[fill] (0,-0.6) circle[radius=0.05];
	\draw[fill] (0.2,-0.6) circle[radius=0.05];
	\draw[fill] (0.4,-0.6) circle[radius=0.05];
}}


\definecolor{darkgreen}{rgb}{0.12549019607843137255,0.4980392156862745098,0.16862745098039215686}

\begin{tikzpicture}
 	[scale=2,node/.style={circle,draw,fill=white,circular drop shadow,thick,inner sep=0pt,minimum size=10mm,fill=white},
 	arrow/.style={->,shorten <=1pt,shorten >=1pt,thick},
 	branch/.style={circle,fill, minimum size=6pt,inner sep=0pt},
 	arrow_d/.style={->,shorten <=1pt,shorten >=1pt,>=stealth',semithick,dashed}]

  
    \pic[rotate=0, color=gray]  (GS) at (-1,0) {gs};
	\node () at (-1.5,0.1)  {VSAT array};

	\draw[arrow] (-1.01,0) -- (0.5,0);
	\node () at (0.55,0)  {$\mathsf{y}$};
	\draw[arrow] (-0.99,0.02) -- (-1.6,-1);
	\node () at (-1.66,-1)  {$\mathsf{x}$};
	\draw[arrow] (-1,-0.01) -- (-1,1.5);
	\node () at (-1,1.55)  {$\mathsf{z}$};

	\draw[thick, darkgreen] (0.45,1.69) -- (-1,0);
	\node (dl) at (0,1.4) {$\color{darkgreen}d_\ell$};

	\draw[darkgreen] (-1.45,1.6) -- (-1,0);
	\node (dl) at (-1.45,1.3) {$\color{darkgreen}d_i$};

	\pic[rotate=5, color=gray] (Satl) at (0.61,2) {sat};
	\node () at (0.5,2.26)  {Satellite $\ell$};
 	
	\pic[rotate=5, color=gray] (Satl) at (-1.5,1.9) {sat};
	\node () at (-1.6,2.18)  {Satellite $i$};

	\draw[dashed, darkgreen] (0,1.15) -- (0,-0.7);
	\draw[dashed, darkgreen] (-1,0) -- (0,-0.7);
 	
	\draw[thick, purple] (-0.25,-0.53) arc (0:50:1.2);
 	\node (th_rx) at (-0.5,-0.2) {$\color{purple}\theta_\ell^\text{el}$};
	\draw[thick, orange] (-1.35,-0.6) arc (240:324:0.7);
	\node (th_rx) at (-1,-0.5) {$\color{orange}\theta_\ell^\text{az}$};
	
%
%

\end{tikzpicture}%
	\caption{Geometric relation between VSAT and satellite $\ell$}
	\label{fig:3D_setup}
\end{figure}
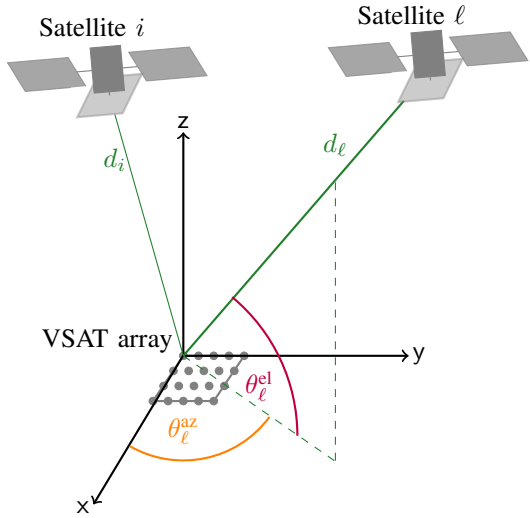

We consider the uplink from a single \ac{tx} to a group of $\NS$ single antenna satellites. The \ac{tx} is equipped with
a \ac{ura} consisting of $\Nt=\Nt^\xdim\Nt^\ydim \ge \NS$ antennas.
Coordinates are stated in a \ac{tx}-centric reference frame as depicted in Fig.~\ref{fig:3D_setup}. In particular, the $\zdim$-axis points toward the \ac{tx}'s zenith, and the $\xdim$- and $\ydim$-axes are aligned with the \ac{tx}'s \ac{ura}. Thus, $\Nt^\xdim$ and $\Nt^\ydim$ are the number of antennas of the \ac{tx} in $\xdim$- and $\ydim$-direction, respectively.
The position of satellite $\ell\in\{1,...,\NS\}$ is given by the triplet  $(d_\ell, \aodl^\text{el}, \aodl^\text{az})$, where $d_\ell$ is the distance between satellite $\ell$ and the \ac{tx} and $\aodl^\text{el}\in[\aod_{\ell,\min}\el, \pi/2]$ and $\aodl^\text{az}\in[0,2\pi]$ are the elevation and azimuth angles, respectively, as shown in Fig.~\ref{fig:3D_setup}. 

The \ac{tx} transmits $\NS$ independent Gaussian data streams $\veks \sim \mathcal{CN}(0,\vekI_{\NS})$, where $\vekI_{\NS}$ is the identity matrix of dimension $\NS$.
These streams are linearly precoded with $\PC=[\pc_1,...,\pc_{\NS}]\in\C^{\Nt\times \NS}$ to obtain the transmit signal $\vekx=\PC\veks$. The \ac{tx} is subject to an average power constraint $\Ptx$, i.e.,
\begin{align}\label{eq:power_constr}
	\Exvl{\vekx^H\vekx}=\tr{\PC\PC^H} \le \Ptx \,.
\end{align}
The received signal $y_\ell$ at satellite $\ell$ is $y_\ell = \vekh_\ell^H \vekx + n_\ell$, where $\vekh_\ell \in \C^{\Nt}$ is the channel from the \ac{tx} to satellite $\ell$ and $n_\ell$ is circularly-symmetric complex white Gaussian noise with power $\sigma_\mathsf{n}^2$.

\subsection{Channel and Error Model} \label{sec:cm}
We assume a \ac{los} path between the \ac{tx} and satellite $\ell$ and the received power from the \ac{nlos} paths to be negligible small.
Then, the elements of $\vekh_\ell$ differ only in their phase, but have the same magnitude \cite{You.Li.Wang.Gao.Xia.Ottersten.2020, Storek.Hofmann.Knopp.2015, Roeper.wcnc.2022}.
Thus, we can define the channel vector $\vekh_\ell$ as a multiplication of a scalar factor $\alpha_\ell\in\C$ and a steering vector $\steeraod_\ell\in\C^{\Nt}$, whose elements have magnitude one, i.e.,
\begin{align}\label{eq:channel}
	\vekh_\ell = \alpha_\ell\steeraod_\ell.
\end{align}
The complex channel gain $\alpha_\ell$ includes the signals attenuation and phase rotation due to the free space propagation and the atmospheric effects.
The steering vector $\steeraod_\ell$ depends on the satellite position.
Due to the regular structure of the \ac{ura}, it is helpful to separate $\steeraod_\ell$ into two steering vectors $\steeraod_\ell^\xdim=[a_{\ell,1}^\xdim,...,a_{\ell,\Nt^\xdim}^\xdim]^T$ and $\steeraod_\ell^\ydim=[a_{\ell,1}^\ydim,...,a_{\ell,\Nt^\ydim}^\ydim]^T$, for the $\xdim$- and $\ydim$-direction, respectively, such that 
\begin{align}\label{eq:steeringvec_kronecker}
	\steeraod_\ell = \steeraod_\ell^\xdim \otimes \steeraod_\ell^\ydim \,,
\end{align}
where $\otimes$ denotes the Kronecker product.
With the \acp{aod} as defined in Fig.~\ref{fig:3D_setup} and given that the distance between the transmit antennas $\DA$ at the \ac{tx} is much smaller than the distance between the \ac{tx} and satellite $\ell$, i.e., $\DA \ll d_\ell$, the elements of the steering vectors are \cite{book:Zooghby.2005}
\begin{subequations}
	\begin{align}
		a_{\ell,m}^\xdim &= e^{-j\nu\DA(m-1)\cos\left(\aodl\el\right)\cos\left(\aodl\az\right)} \,, &&m=1,...,\Nt^\xdim \,,
		\\
		a_{\ell,n}^\ydim &= e^{-j\nu\DA(n-1)\cos\left(\aodl\el\right)\sin\left(\aodl\az\right)} \,, &&n=1,...,\Nt^\ydim \,,
	\end{align}
\end{subequations}
where $\nu$ is the wavenumber of the carrier wave .
Furthermore, we define the space angles $\phi_\ell^\xdim=\cos\left(\aodl\el\right)\cos\left(\aodl\az\right)$ and $\phi_\ell^\ydim=\cos\left(\aodl\el\right)\sin\left(\aodl\az\right)$.
While the \acp{aod} $\aodl\el$ and $\aodl\az$ are related to the spherical coordinates, the space angles $\phi_\ell^\xdim$ and $\phi_\ell^\ydim$ are related to the Cartesian \xdim- and \ydim- coordinates of satellite $\ell$, respectively.
Note that in practical systems only imperfect knowledge of the satellites positions can be assumed.
Let $\erraod_\ell^\xdim$ and $\erraod_\ell^\ydim$ be the estimation errors for the space angles in \xdim- and \ydim-direction, respectively, the estimated space angles are
\begin{align}
	\hat{\phi}_{\ell}^\xdim 
	&= \phi_{\ell}^{\mathsf{x}} + \erraod_\ell^\xdim, \qquad
	\hat{\phi}_{\ell}^{\mathsf{y}} 
	= \phi_{\ell}^{\mathsf{y}} + \erraod_\ell^\ydim \,.
\end{align}
Then, the $k=n+(m-1)\Nt^\ydim$th element of the channel vector $\vekh_\ell$ can be written as
\begin{align}
	\begin{split}
		h_{\ell,k} &= \alpha_\ell a_{\ell,m}^\xdim a_{\ell,n}^\ydim \\
		&= \alpha_\ell e^{-j\nu\DA\left((m-1)\left(\hat{\phi}_{\ell}^\xdim - \erraod_\ell^\xdim\right) +  (n-1)\left(\hat{\phi}_{\ell}^\ydim - \erraod_\ell^\ydim\right)\right)} \,,
	\end{split}
\end{align}
which includes three statistically independent random variables from the transmitters perspective, i.e., $\alpha_\ell$, $\erraod_\ell^\xdim$ and $\erraod_\ell^\ydim$. 
The complex channel gain $\alpha_\ell$ is a circularly-symmetric random variable with variance $\E\{|\alpha_\ell|^2\} = \sigma_{\alpha_\ell}^2$. Note that the variance $\sigma_{\alpha_\ell}^2$ can be different for each satellite $\ell$, due to the different path losses.
For the estimation errors of the space angles $\erraod_\ell^\xdim$ and $\erraod_\ell^\ydim$, we assume the same probability distribution for every satellite $\ell$ and for both directions.
Their probability distribution is described by the characteristic function $\cfaod(t)$ \cite{book:Kobayashi.Mark.Turin.2011}.

\subsection{Upper bound on the channel capacity}
To obtain an upper bound on the throughput performance, perfect \ac{csi} at the transmitter and receiver as well as instantaneous and error-free inter-satellite communication is assumed. Hence, perfect coordination between satellites is possible and the system can be modelled 
as an $\Nt\times\NS$ \ac{mimo} system. 
The joint receive signal of all satellites is
	$\veky = \vekH \vekx + \vekn$,
with the channel matrix $\vekH=[\vekh_1,...,\vekh_{\NS}]^H$ and the noise vector $\vekn=[n_1,...,n_{\NS}]^T$, where the components of $\vekn$ are mutually independent. 
Let  $\lambda_\mu$ be the $\mu$th eigenvalue of $\vekH\vekH^H$, the capacity of this channel is \cite{Telatar.1999}
	\begin{align}
		C 
		&= \sum_{\mu=1}^{\NS} \log_2\left(1 + \lambda_\mu\frac{p_\mu}{\nvar} \right), \label{eq:rate_opt}
	\end{align}
where and $p_\mu$ is the optimal power allocated to the $\mu$th stream. Given the power constraint $\sum_\mu p_\mu = \Ptx$, the optimal power allocation is obtained from the water-filling algorithm \cite{Telatar.1999}.

\section{Proposed Uplink Approach}\label{sec:pc}
In this section, we present a novel low-complexity robust precoder for uplink transmission from a \ac{tx} to multiple satellites based on imperfect knowledge of the satellites positions. For sufficiently large distances between the satellites, the channel vectors become nearly orthogonal, i.e., $\steeraod_i^H\steeraod_\ell\approx 0$. In this case, the channel capacity is maximized, because the eigenvalues of $\vekH\vekH^H$ are almost equal \cite{Roeper.wcnc.2022}.
Given orthogonal channels, the capacity can be achieved by transmitting independent data streams \cite{Telatar.1999}.\footnote{Observe that this does not imply the transmission of several independent messages. See, e.g., \cite{Ariyavisitakul2000} or \cite[Sec.~5.8.1]{Heath2019}.} Thus, we assign one stream per satellite and then design the precoders.
Given independent data streams, no inter-satellite communication is required for estimation and decoding because satellite $\ell$ can estimate and decode the transmitted symbol $s_\ell$ independently from the other satellites.
Numerical results in Section \ref{sec:sim_distance} show that this approach is, despite its simplicity, capacity achieving for sufficiently large orbital separations and perfect position knowledge of the satellites.
The received signal at satellite $\ell$ is
	$y_\ell = \vekh_\ell^H\pc_\ell s_\ell + \sum_{i\ne \ell} \vekh_\ell^H\pc_i s_i + n_\ell$.
Correspondingly, the instantaneous \ac{sinr} $\sinr_\ell$ at satellite $\ell$ is given as
\begin{align} \label{eq:sinrl}
	\sinr_\ell = \frac{\left|\hugu\right|^2}{\sum_{i\neq \ell}|\hugv|^2 + \nvar}
\end{align}
and the achievable rate $R$ is equal to the sum rate
\begin{align}\label{eq:sum_rate}
	R = \sum_{\ell=1}^{\NS} \log_2\left(1 + \sinr_\ell\right) \,.
\end{align}
The goal is to obtain precoders that maximize this rate. However, the exact values of $\vekh_\ell$ are not known but only their estimates based on $\hat\phi_\ell^\xdim$ and $\hat\phi_\ell^\ydim$. Furthermore,
directly maximizing this sum rate is NP-hard and has, to the best of our knowledge, no easy analytical solution. In the following, we solve a slightly simplified version of this problem that, as will be seen later in Sec.~\ref{sec:simulation}, will result in a solution that is often close to the optimal throughput performance.

\subsection{Problem Formulation}
Observe that the \acp{sinr} $\sinr_\ell$ in \eqref{eq:sinrl} are coupled through their denominator.
Substituting $\sinr_\ell$ in \eqref{eq:sum_rate} with the instantaneous \ac{slnr} for satellite $\ell$ defined as
\begin{equation}
\gamma_\ell = \frac{|\hugu|^2}{\sum_{i\ne \ell}|\hvgu|^2 + \nvar}
\end{equation}
and assuming equal power allocation among streams, leads to the substitute optimization problem
\begin{align}
		\max_{\pc_\ell}\enskip \frac{|\hugu|^2}{\sum_{i\ne \ell}|\hvgu|^2 + |\nvar} \quad
		\text{s.t.} \quad \pc_\ell^H\pc_\ell \le \frac{\Ptx}{\NS},
\end{align}
which has to be solved for each $\ell \in \{1,...,\NS\}$.
However, this problem still relies on the exact knowledge of the channels. Our goal is to design robust precoders that take the statistics of the estimation error into account. This can be achieved by optimizing over the mean \ac{slnr} $\slnr=\Exvl{\gamma}$, i.e.,
\begin{align}\label{eq:opt_problem}
		\max_{\pc_\ell} \Exvl{\frac{|\hugu|^2}{\sum_{i\ne \ell}|\hvgu|^2 + \nvar}} \quad
		\text{s.t.} \quad \pc_\ell^H\pc_\ell \le \frac{\Ptx}{\NS} \,,
\end{align}
where the expectation is taken \ac{wrt} the channel vectors $\vekh_1, \dots, \vekh_{\NS}$. Then, the robust precoders $\pc_\ell^{\text{rob}}$ are a solution of \eqref{eq:opt_problem}.

\subsection{Robust Precoding via CF}
The solution of \eqref{eq:opt_problem} requires a closed-form expression of the autocorrelation matrix of the channel $\Exvl{\vekh_\ell\vekh_\ell^H}$, which is closely related to the autocorrelation matrix of the steering vectors.
In particular, let
$\corrmtxl=\Exvl{\steeraod_\ell\steeraod_\ell^H}$ be the  autocorrelation matrix of the steering vector $\steeraod_\ell$.
With \eqref{eq:steeringvec_kronecker} and due to the statistical independence of the position uncertainties $\erraod_\ell^\xdim$ and $\erraod_\ell^\ydim$, we can write
\begin{subequations}\label{eq:corrmtx_steering_vec}
	\begin{align}
		\corrmtxl &= \Exvl{\steeraod_\ell\steeraod_\ell^H} = \Exvl{\left(\steeraod_\ell^\xdim \otimes \steeraod_\ell^\ydim\right)\left(\steeraod_\ell^\xdim \otimes \steeraod_\ell^\ydim\right)^H} \\
		&= \Exvl{\steeraod_\ell^\xdim{\steeraod_\ell^\xdim}^H} \otimes \Exvl{\steeraod_\ell^\ydim{\steeraod_\ell^\ydim}^H} = \corrmtxlx \otimes \corrmtxly .
	\end{align}
\end{subequations}
The random variables in $\corrmtxlx$ and $\corrmtxly$ are only the estimation errors $\erraod_\ell^\xdim$ and $\erraod_\ell^\ydim$, respectively. The characteristic function $\cfaod(t)$ of these random variables is defined as \cite{book:Kobayashi.Mark.Turin.2011}
\begin{align}\label{eq:cf_def}
	\cfaod(t) = \Exvl{e^{jt\erraod_\ell^\xdim}} = \Exvl{e^{jt\erraod_\ell^\ydim}}.
\end{align}
Thus, the $(m,m')$th element of the autocorrelation matrix $\corrmtxlx$ is given by
\begin{subequations}
\begin{align}
	\left[\corrmtxlx\right]_{m,m'} &= \Exvl{e^{-j\nu\DA(m-m')\left(\hat{\phi}_{\ell}^\xdim - \erraod_\ell^\xdim \right)}} \\
	&= e^{-j\nu\DA(m-m')\hat{\phi}_\ell} \cfaod\!\left(\nu\DA(m'-m)\right).
\end{align}
\end{subequations}
The elements of $\corrmtxly$ are given, analogously.

Now, we can formulate the optimal precoder, as stated in the following theorem.
\begin{theorem}\label{theorm:pc_robust}
    The robust precoder, i.e., the solution to \eqref{eq:opt_problem}, is
\begin{align}\label{eq:pc_rob}
	\pc_\ell^\mathrm{rob} = \frac{\Ptx}{\NS}\, \eigvec_{\ell,\max} \,,
\end{align}
where $\eigvec_{\ell,\max}$ is the eigenvector corresponding to the largest eigenvalue of ${(\sum_{i\ne \ell} \sigma_{\alpha_i}^2 \corrmtxi + \NS\nvar/\Ptx\vekI_{\Nt} )^{-1}\sigma_{\alpha_\ell}^2\corrmtxl}$.
\end{theorem}

\begin{IEEEproof}
Note that $\alpha_\ell$, $\erraod_\ell^\xdim$ and $\erraod_\ell^\ydim$ are statistically independent of $\alpha_i$,  $\erraod_i^\xdim$ and $\erraod_i^\ydim$ for $i\ne\ell$. Therefore, $\vekh_\ell$ and $\vekh_i$ are statistically independent, too. Thus, we can rewrite the objective function \eqref{eq:opt_problem} as
\begin{subequations}
	\begin{align}
		\slnr_\ell &= \Exvl{\frac{|\hugu|^2}{\sum_{i\ne \ell}|\hvgu|^2 + \nvar}} \\
		{} \nonumber \\
		&= \frac{\pc_\ell^H\Exvl{\vekh_\ell\vekh_\ell^H}\pc_\ell }{\pc_\ell^H\left(\sum_{i\ne \ell}\Exvl{\vekh_i\vekh_i^H} + \frac{\nvar}{p_\ell}\vekI_{\Nt} \right)\pc_\ell} \label{eq:slnr_rayleigh}
	\end{align}
\end{subequations}
    where $p_\ell = \pc_\ell^H\pc_\ell$ is the transmit power of the $\ell$th stream.
    
    With \eqref{eq:channel} and \eqref{eq:corrmtx_steering_vec}, the expected value in \eqref{eq:slnr_rayleigh} can be written as
	    $\Exvl{\vekh_\ell\vekh_\ell^H} = \sigma_{\alpha_\ell}^2\corrmtxl$. 
    Furthermore, the term in \eqref{eq:slnr_rayleigh} is monotonically increasing with the transmit power $p_\ell$. Therefore, at the optimum solution the constraint must be active, i.e., $p_\ell = \Ptx/\NS$.
    Correspondingly, the optimization problem \eqref{eq:opt_problem} is equivalent to
    \begin{subequations}\label{eq:opt_problem_equiv}
	\begin{align}
		&\max_{\pc_\ell} \frac{\sigma_{\alpha_\ell}^2\pc_\ell^H\corrmtxl\pc_\ell }{\pc_\ell^H\left(\sum_{i\ne \ell}\sigma_{\alpha_i}^2\corrmtxi + \frac{\NS\nvar}{\Ptx}\vekI \right)\pc_\ell} \label{eq:objective_equiv} \\
		&\text{s.t.} \quad \pc_\ell^H\pc_\ell = \frac{\Ptx}{\NS} \label{eq:constraint_equiv}
	\end{align}
    \end{subequations}
    The objective function \eqref{eq:objective_equiv} is a generalized Rayleigh quotient. Consequently, the maximum is achieved by the scaled eigenvector corresponding to the largest eigenvalue of $(\sum_{i\ne \ell}\sigma_{\alpha_i}^2\corrmtxi + \NS\nvar/\Ptx\vekI_{\Nt} )^{-1}\corrmtxl$ \cite{Sadek.Tarighat.Sayed.2007}, i.e., the precoder $\pc_\ell^{\mathrm{rob}}$ must be proportional to $\eigvec_{\ell,\max}$. Given the constraint \eqref{eq:constraint_equiv}, the only solution is given by \eqref{eq:pc_rob}.
\end{IEEEproof}

The precoder in Theorem \ref{theorm:pc_robust} optimizes \eqref{eq:opt_problem} for any probability distribution of $\erraod_\ell^\xdim$ and $\erraod_\ell^\ydim$. For the special case of perfect position knowledge, we obtain a closed-form solution, as stated in the following corollary.

\begin{corollary}\label{theorm:pc_perfect}
    For perfect knowledge of the steering vectors, the optimal precoder for \eqref{eq:opt_problem} is 
\begin{align}\label{eq:pc_per}
	\pc_\ell^{\mathrm{per}} = \beta\left(\sum_{i=1}^{\NS} \sigma_{\alpha_i}^2 \steeraod_i\steeraod_i^H + \frac{\NS\nvar}{\Ptx}\vekI_{\Nt} \right)^{-1} \steeraod_\ell
\end{align}
where 
the normalization coefficient $\beta$ is chosen such that $\tr{\PC_\ell\PC_\ell^H}=\Ptx/\NS$.
\end{corollary}

\begin{IEEEproof}
     For perfect position knowledge, i.e., ${\erraod_\ell^\xdim=\erraod_\ell^\ydim=0}$, the characteristic function $\cfaod(t)$ becomes one. Thus, the objective function is
     \begin{align}
         \slnr|_{\cfaod(t)=1} = \frac{\sigma_{\alpha_\ell}^2 \pc_\ell^H\steeraod_\ell\steeraod_\ell^H\pc_\ell }{\pc_\ell^H\left(\sum_{i\ne \ell} \sigma_{\alpha_i}^2 \steeraod_i\steeraod_i^H + \frac{\NS\nvar}{\Ptx}\vekI \right)\pc_\ell} .
     \end{align}
     Consequently, the optimal precoder with perfect position knowledge must be a scaled eigenvector of $\left(\sum_{i\ne \ell}\sigma_{\alpha_i}^2\steeraod_i\steeraod_i^H + \NS\nvar/\Ptx\vekI_{\Nt} \right)^{-1}\steeraod_\ell\steeraod_\ell^H$. Such an eigenvector is given by \eqref{eq:pc_per} \cite{Patcharamaneepakorn.Armour.Doufexi.2012}
\end{IEEEproof}

Furthermore, given perfect position knowledge and orthogonal channels, the following corollary holds as well.

\begin{corollary}\label{cor:optimality}
    If $\forall i\ne\ell$, the steering vectors $\steeraod_i$ and $\steeraod_\ell$ are orthogonal and the path losses $|\alpha_i|^2$ and $|\alpha_\ell|^2$ are the same, i.e., $\steeraod_i^H\steeraod_\ell=0$ and $|\alpha_i|^2=|\alpha_\ell|^2$, respectively, the precoder with perfect position knowledge \eqref{eq:pc_per} is capacity achieving.
\end{corollary}

\begin{IEEEproof}
	Note that the precoder \eqref{eq:pc_per} is capacity achieving, if its columns are given by the right singular vectors of $\vekH$ and the norm of each column is equal to the optimal power allocation, obtained via the water-filling algorithm \cite{Telatar.1999}.
	Let $\Steeraod=[\steeraod_1,...,\steeraod_{\NS}]$, with $\Steeraod^H\Steeraod=\Nt\vekI_{\NS}$, and $|\alpha_i|^2=|\alpha_\ell|^2=|\alpha|^2$, for all $i$ and $\ell$, the precoder matrix $\PC^{\mathrm{per}}=[\pc_1^{\mathrm{per}},...,\pc_{\NS}^{\mathrm{per}}]$ can be written as
	\begin{subequations}
		\begin{align}
			\PC^{\mathrm{per}} &= \beta  \left(|\alpha|^2\Steeraod\Steeraod^H + \frac{\NS\nvar}{\Ptx}\vekI_{\Nt}\right)^{-1} \Steeraod\\
			&= \beta \Steeraod \left(|\alpha|^2\Steeraod^H\Steeraod + \frac{\NS\nvar}{\Ptx}\vekI_{\NS}\right)^{-1}  \label{eq:pc_mtx} \\
			&= \sqrt{\Ptx/\Nt}\Steeraod \,, \label{eq:pc_mtx_orth}
		\end{align}
	\end{subequations}
	where \eqref{eq:pc_mtx} follows from the the matrix inversion lemma \cite{book:Dietrich.2008} and \eqref{eq:pc_mtx_orth} is obtained due to the orthogonal steering vectors.
    
    Now, given the channel model \eqref{eq:channel}, the matrix $\vekH$ can be factorized via the \ac{svd} as
    \begin{align}
    	\vekH = \left(\diag{\alpha_1,...,\alpha_{\NS}}\Steeraod \right)^H = \vekU\boldsymbol{\Sigma}\vekV^H \,,
    \end{align}
	where $\vekU=1/|\alpha|\diag{\alpha_1,...,\alpha_{\NS}}^H$ and $\vekV=1/\sqrt{\Nt} [\Steeraod, \vekv_{\NS+1},...,\vekv_{\Nt} ]$ are unitary matrices.
	The vectors $\{\vekv_{\NS+1},...,\vekv_{\Nt}\}$ are the right singular vectors belonging to the nullspace of the channel matrix $\vekH$. 
	Furthermore, $\boldsymbol{\Sigma} = [|\alpha|\sqrt{\Nt}\vekI_{\NS}, \nullvec_{\NS\times(\Nt-\NS)}]$, where $\nullvec_{\NS\times(\Nt-\NS)}$ is an all zero matrix of dimension $\NS\times(\Nt-\NS)$, is a rectangular diagonal matrix with the singular values of $\vekH$ on its diagonal.
	Thus, the precoder matrix \eqref{eq:pc_mtx_orth} is proportional to the right singular vectors of $\vekH$, which do not belong to the nullspace.
	Finally, as there is only a single non-zero singular value with multiplicty $\NS$, the capacity is achieved by allocating the same power to each right singular vector. This is given by the precoder $\PC^\text{per}$, which concludes the proof.
\end{IEEEproof}

Note that in Corollary~\ref{cor:optimality}, the assumptions on the channel are very strict. In the next section, we show numerically that the precoder also achieves the capacity in more general cases.

\section{Numerical Evaluations}\label{sec:simulation}
To evaluate the proposed precoder approach, we consider a satellite swarm in triangle formation with a fixed inter-satellite distance $\DS$ between any of these $\NS=3$ satellites. The \ac{tx} is equipped with a $32\times32$ \ac{ura} with an antenna spacing of $\DA=\SI{2.5}{cm}$. The antenna gain at the \ac{tx} and the satellites are $\GTx=\SI{43.2}{dBi}-10\log_{10}(\Nt)\approx\SI{13.1}{dBi}$ and  $\GRx=\SI{30.5}{dBi}-10\log_{10}(\NS)\approx\SI{25.7}{dBi}$, respectively, to match the 3GPP recommendation \cite{3GPP.TR.38.821}. The carrier frequency and the noise power are $\fc = \SI{30}{\giga\hertz}$ and $P_{\text{N}}=\SI{-120}{dBW}$, respectively.
The channel is modeled as a pure \ac{los} channel. Thus, the scaling factor is $\alpha_\ell=1/L_\ell e^{j\phi_{\ell,0}}$, where $\phi_{\ell,0}\sim \mathcal{U}(0,2\pi)$ is a random phase rotation and $L_\ell$ is the path loss, including the antenna gains $\GTx$ and $\GRx$, free space path loss, shadow fading, clutter loss \cite{3GPP.TR.38.811}, atmospheric gas absorption \cite{ITUR.P.676} and tropospheric scintillation \cite{ITUR.P.618,ITUR.P.531}.
Furthermore, the altitude of the satellites is $d_0=\SI{600}{km}$ and the minimum elevation angle is $\aod_{\ell,\min}^{\el}=30^\circ$.

\subsection{Optimal Inter-Satellite Distance}\label{sec:sim_distance}
\begin{figure}
%
%
\definecolor{mycolor1}{rgb}{0.00000,0.44700,0.74100}%
\definecolor{mycolor2}{rgb}{0.49400,0.18400,0.55600}%
\begin{tikzpicture}

\begin{axis}[%
width=0.85\columnwidth,
height=1.5in,
scale only axis,
xmode=log,
xmin=0.1,
xmax=10000,
xminorticks=true,
xlabel style={font=\color{white!15!black}},
xlabel={Inter-satellite  distance $D_\text{S}\,$km},
ymin=1,
ymax=10,
ylabel style={font=\color{white!15!black}},
ylabel={Achievable Rate [bps/Hz]},
axis background/.style={fill=white},
xmajorgrids,
xminorgrids,
ymajorgrids,
legend style={at={(0.01,0.98)}, anchor=north west, legend cell align=left, align=left, fill opacity=0.8, draw opacity=1, text opacity=1, draw=white!80!black}
]
\addplot [color=mycolor1, thick]
  table[row sep=crcr]{%
0.1	6.36794286115644\\
0.123548288825675	6.36757630640915\\
0.152641796717523	6.37345137509696\\
0.188586327877265	6.37011297942412\\
0.232995181051537	6.36347755705494\\
0.287861559235457	6.3662808166369\\
0.355648030622313	6.37130235439254\\
0.439397056076079	6.37166604168044\\
0.542867543932386	6.36747320936247\\
0.670703561118431	6.36412498322916\\
0.828642772854684	6.35904244478856\\
1.02377396633958	6.37053553006194\\
1.2648552168553	6.36941260501743\\
1.562706976547	6.40211345936466\\
1.93069772888325	6.4974616396697\\
2.38534400643142	6.63078054724355\\
2.94705170255181	6.83667893570735\\
3.64103194931068	7.13644124250885\\
4.49843266896945	7.51688125172591\\
5.55773658648688	7.9017631084446\\
6.866488450043	8.35596165939489\\
8.48342898244072	8.76527166098523\\
10.4811313415469	9.11169036722723\\
12.9492584220526	9.32332952758179\\
15.9985871960606	9.45759442771237\\
19.7659807170163	9.57932384764365\\
24.4205309454865	9.60324029431812\\
30.171148105293	9.62948211272438\\
37.2759372031494	9.64435672506641\\
46.0537825582242	9.65590421381985\\
56.898660290183	9.65789729684516\\
70.2973211532548	9.66141255470949\\
86.8511373751353	9.64768864192989\\
107.303094052616	9.63537533521216\\
132.571136559011	9.64767777945089\\
163.789370695406	9.63837224835339\\
202.358964772516	9.62956842236296\\
250.011038261793	9.58802392838959\\
308.884359647748	9.57673512794422\\
381.621340794936	9.51211531539307\\
471.486636345739	9.4490551068482\\
582.513671246893	9.3287269342613\\
719.685673001152	9.18589282604043\\
889.159333916464	8.95028632765694\\
1098.54114198756	8.622681890577\\
1357.22878297165	8.16712526038211\\
1676.83293681101	7.57394485437918\\
2071.69839989531	6.78425750173056\\
2559.54792269953	5.86294077657275\\
3162.27766016838	4.81679795547923\\
};
\addlegendentry{$C$}

\addplot [color=mycolor2, thick]
  table[row sep=crcr]{%
0.1	1.65286165615954\\
0.123548288825675	1.65459278398711\\
0.152641796717523	1.65722976629115\\
0.188586327877265	1.66074840089655\\
0.232995181051537	1.66634718639793\\
0.287861559235457	1.67603302224645\\
0.355648030622313	1.69095694606194\\
0.439397056076079	1.71471009109622\\
0.542867543932386	1.7514279346672\\
0.670703561118431	1.80833208178072\\
0.828642772854684	1.892082053084\\
1.02377396633958	2.02380833852768\\
1.2648552168553	2.2039210144998\\
1.562706976547	2.46109314070474\\
1.93069772888325	2.81877356095878\\
2.38534400643142	3.2620238935755\\
2.94705170255181	3.81325478374358\\
3.64103194931068	4.48201639450569\\
4.49843266896945	5.25030183818044\\
5.55773658648688	6.01264781455493\\
6.866488450043	6.85924552689344\\
8.48342898244072	7.63838010906194\\
10.4811313415469	8.32089551667148\\
12.9492584220526	8.79162114413784\\
15.9985871960606	9.13169308022056\\
19.7659807170163	9.42791991993725\\
24.4205309454865	9.54422908671407\\
30.171148105293	9.60230768397154\\
37.2759372031494	9.61935525151689\\
46.0537825582242	9.63699190617248\\
56.898660290183	9.64484051605896\\
70.2973211532548	9.64932540059738\\
86.8511373751353	9.63661813432195\\
107.303094052616	9.62458961577259\\
132.571136559011	9.63717683339673\\
163.789370695406	9.62789995914924\\
202.358964772516	9.61921497249365\\
250.011038261793	9.57732309009504\\
308.884359647748	9.56571605431542\\
381.621340794936	9.49621808346498\\
471.486636345739	9.43254674706411\\
582.513671246893	9.3101825435631\\
719.685673001152	9.163228981453\\
889.159333916464	8.92605122146541\\
1098.54114198756	8.59743385986162\\
1357.22878297165	8.13569322211577\\
1676.83293681101	7.5340883949799\\
2071.69839989531	6.72977642126957\\
2559.54792269953	5.79236007481221\\
3162.27766016838	4.71157931593362\\
};
\addlegendentry{$R$}

\addplot [color=red, dashed, thick]
  table[row sep=crcr]{%
40 10\\
40 0\\
};
\node[] at (axis cs: 80,6.5) {$\text{D}_{\text{S,opt}}$};

\end{axis}
\end{tikzpicture}%
	\caption{Achievable rate for different inter-satellite distances $\DS$ and perfect position knowledge}
	\label{fig:RvsDs}
\end{figure}
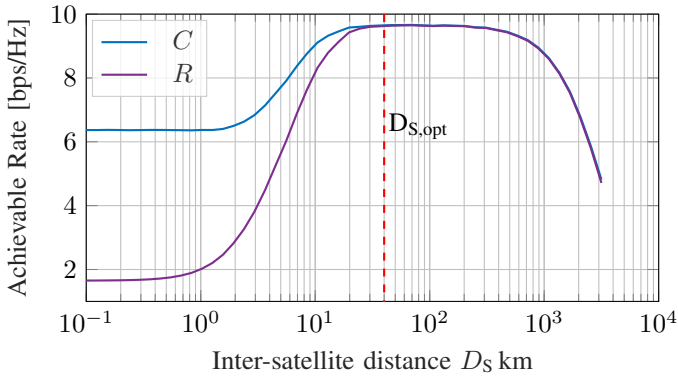

In \cite{Roeper.wcnc.2022}, the optimal inter-satellite distance for a simplified downlink scenario has been derived. In this subsection, we evaluate the performance \ac{wrt} the inter-satellite distance for an uplink scenario and a fixed transmit power of $\Ptx=\SI{5}{dBW}$.
Given the altitude $d_0=\SI{600}{km}$, the minimum elevation angle $\aod_{\ell,\min}^{\el}=30^\circ$, $\Nt^\xdim=32$ antennas along the \xdim-axis and $\nu\DA=5\pi$, the analytic solution for the optimum inter-satellite distance is $D_{\text{S,opt}}\approx \SI{40}{km}$ \cite{Roeper.wcnc.2022}.
In Fig.~\ref{fig:RvsDs}, the channel capacity $C$ and the sum rate $R$ with the proposed precoder and perfect position knowledge are shown. It can be observed that both rates increases with increasing inter-satellite distances $\DS$ up to a certain distance and then slightly decrease. The inter-satellite distance, where both rates achieve their maximum matches with the analytic solution $D_{\text{S,opt}}$ from \cite{Roeper.wcnc.2022}. For $\DS<D_{\text{S,opt}}$, the sum rate is significantly smaller than the channel capacity, due to the relatively big difference between the maximum and minimum eigenvalue. If $\DS\ge D_{\text{S,opt}}$, all eigenvalues are approximately the same because $\steeraod_i^H\steeraod_\ell\approx 0$, for $i\ne\ell$. Therefore, the difference between the sum rate and the capacity is negligible small. The performance degradation for very large inter-satellite distances is due to the increased path loss averaged over the satellites.



\subsection{Robust Precoding}\label{sec:sim_robust}
Now, we evaluate the performance of the proposed robust precoder \eqref{eq:pc_rob} for a constant inter-satellite distance $\DS=\SI{40}{km}$. Furthermore, we assume two different error distributions of the position uncertainty.
The robust precoder requires knowledge about the \ac{cf} of the probability distribution.
Therefore, the \acp{cf} of both probability distribution are given in the appendix.
For comparison, a heuristic precoder $\pc_\ell^\text{heu}$ for imperfect knowledge is obtained by substituting the true steering vectors by the estimated ones in \eqref{eq:pc_per}, i.e.,
\begin{align}\label{eq:pc_heu}
	\pc_\ell^{\text{heu}} = \beta\left(\sum_{i\ne\ell} \sigma_{\alpha_i}^2 \steeraodest_i\steeraodest_i^H + \frac{\NS\nvar}{\Ptx}\vekI_{\Nt} \right)^{-1} \steeraodest_\ell .
\end{align}
Note that for perfect position knowledge, the heuristic precoder \eqref{eq:pc_heu} as well as the robust precoder \eqref{eq:pc_rob} are the same.
 
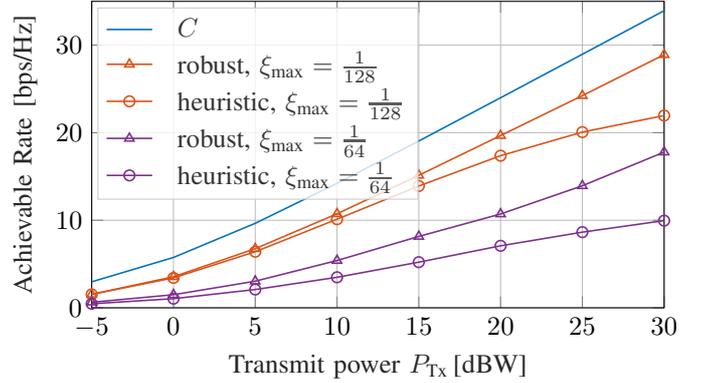
\begin{figure}
%
%
\definecolor{mycolor1}{rgb}{0.00000,0.44700,0.74100}%
\definecolor{mycolor2}{rgb}{0.92900,0.69400,0.12500}%
\definecolor{mycolor3}{rgb}{0.49400,0.18400,0.55600}%
\definecolor{mycolor4}{rgb}{0.85098,0.32549,0.09804}%
\definecolor{mycolor5}{rgb}{0.00000,0.44706,0.74118}%
\begin{tikzpicture}

\begin{axis}[%
width=0.85\columnwidth,
height=1.6in,
scale only axis,
xmin=-5,
xmax=30,
xlabel style={font=\color{white!15!black}},
xlabel={Transmit power $P_{\text{Tx}}\,$[dBW]},
ymin=0,
ymax=35,
ylabel style={font=\color{white!15!black}},
ylabel={Achievable Rate [bps/Hz]},
axis background/.style={fill=white},
title style={font=\bfseries},
xmajorgrids,
ymajorgrids,
legend style={at={(0.01,0.98)}, anchor=north west, legend cell align=left, align=left, fill opacity=0.8, draw opacity=1, text opacity=1, draw=white!80!black}
]
\addplot [color=mycolor1, semithick]
  table[row sep=crcr]{%
-5	2.96923624664424\\
0	5.750262398618\\
5	9.6516726759071\\
10	14.2138541868869\\
15	19.0525449330347\\
20	23.9885566926479\\
25	28.9564892692588\\
30	33.9346371532612\\
};
\addlegendentry{$C$}

\addplot [color=mycolor4, semithick, mark=triangle, mark options={solid}]
  table[row sep=crcr]{%
-5	1.54207229265627\\
0	3.54647416828085\\
5	6.75414262648155\\
10	10.7349849446074\\
15	15.1353760512612\\
20	19.6710827956976\\
25	24.2311882451033\\
30	28.9084901401494\\
};
\addlegendentry{robust, $\xi_{\text{max}}=\frac{1}{128}$}

\addplot [color=mycolor4, semithick, mark=o, mark options={solid}]
  table[row sep=crcr]{%
-5	1.53641771064961\\
0	3.41853866749069\\
5	6.40389679512302\\
10	10.1198129668915\\
15	13.9122002102521\\
20	17.3714226239475\\
25	20.0671667789226\\
30	21.957308579385\\
};
\addlegendentry{heuristic, $\xi_{\text{max}}=\frac{1}{128}$}

\addplot [color=mycolor3, semithick, mark=triangle, mark options={solid}]
  table[row sep=crcr]{%
-5	0.635828801997173\\
0	1.49615136687272\\
5	3.04046474322284\\
10	5.40012282900025\\
15	8.15070620180786\\
20	10.7145617462995\\
25	13.9378991798881\\
30	17.7847145743487\\
};
\addlegendentry{robust, $\xi_{\text{max}}=\frac{1}{64}$}

\addplot [color=mycolor3, semithick, mark=o, mark options={solid}]
  table[row sep=crcr]{%
-5	0.462308218919432\\
0	1.0426553610168\\
5	2.08921158758081\\
10	3.4851937039054\\
15	5.21347647847765\\
20	7.08115153395656\\
25	8.6400783657842\\
30	9.95496824895983\\
};
\addlegendentry{heuristic, $\xi_{\text{max}}=\frac{1}{64}$}

\end{axis}

\end{tikzpicture}%
	\caption{Achievable rate of robust and heuristic precoder in case of uniformly distributed position uncertainty}
	\label{fig:robust_uniform}
\end{figure}

In Fig.~\ref{fig:robust_uniform}, the achievable rates for the proposed robust and heuristic precoder \eqref{eq:pc_rob} and \eqref{eq:pc_heu}, respectively, are shown for a uniformly distributed position uncertainty. Thus, the error of the space angles are uniformly distributed, i.e., $\erraod_\ell^\xdim,\erraod_\ell^\ydim\sim\mathcal{U}(-\erraod_{\max},\erraod_{\max})$.
It can be seen, that the sum rate with the heuristic precoder degrades, especially for high transmit powers. For the robust precoder, the sum rate is almost parallel to the channel capacity. Thus, the performance degradation is less severe and the robust precoder clearly outperforms the heuristic precoder.

\begin{figure}
%
%
\definecolor{mycolor1}{rgb}{0.00000,0.44700,0.74100}%
\definecolor{mycolor2}{rgb}{0.92900,0.69400,0.12500}%
\definecolor{mycolor3}{rgb}{0.49400,0.18400,0.55600}%
\definecolor{mycolor4}{rgb}{0.85098,0.32549,0.09804}%
\definecolor{mycolor5}{rgb}{0.00000,0.44706,0.74118}%
\begin{tikzpicture}

\begin{axis}[%
width=0.85\columnwidth,
height=1.6in,
scale only axis,
xmin=-5,
xmax=30,
xlabel style={font=\color{white!15!black}},
xlabel={Transmit power $P_{\text{Tx}}\,$[dBW]},
ymin=0,
ymax=35,
ylabel style={font=\color{white!15!black}},
ylabel={Achievable Rate [bps/Hz]},
axis background/.style={fill=white},
title style={font=\bfseries},
xmajorgrids,
ymajorgrids,
legend style={at={(0.01,0.98)}, anchor=north west, legend cell align=left, align=left, fill opacity=0.8, draw opacity=1, text opacity=1, draw=white!80!black}
]
\addplot [color=mycolor1, semithick]
  table[row sep=crcr]{%
-5	2.96923624664424\\
0	5.750262398618\\
5	9.6516726759071\\
10	14.2138541868869\\
15	19.0525449330347\\
20	23.9885566926479\\
25	28.9564892692588\\
30	33.9346371532612\\
};
\addlegendentry{$C$}

\addplot [color=mycolor4, semithick, mark=triangle, mark options={solid}]
  table[row sep=crcr]{%
-5	1.67698491490508\\
0	3.7808458529783\\
5	6.98141909412452\\
10	10.9696298247813\\
15	15.259238097284\\
20	19.597713359857\\
25	23.9406521306419\\
30	27.9955378166219\\
};
\addlegendentry{robust, $\sigma_{\xi}^2=2\cdot 10^{-5}$}

\addplot [color=mycolor4, semithick, mark=o, mark options={solid}]
  table[row sep=crcr]{%
-5	1.65620742268192\\
0	3.60121370249274\\
5	6.54020255534988\\
10	10.2335677023414\\
15	13.9991630183999\\
20	17.4963097694643\\
25	20.2652619841782\\
30	22.2411285423933\\
};
\addlegendentry{heuristic, $\sigma_{\xi}^2=2\cdot 10^{-5}$}

\addplot [color=mycolor3, semithick, mark=triangle, mark options={solid}]
  table[row sep=crcr]{%
-5	0.7932484012453\\
0	1.83983160700658\\
5	3.43504278843551\\
10	5.79511897852653\\
15	8.2150566541454\\
20	10.6309646761543\\
25	12.492347339904\\
30	13.8339411483549\\
};
\addlegendentry{robust, $\sigma_{\xi}^2=8\cdot 10^{-5}$}

\addplot [color=mycolor3, semithick, mark=o, mark options={solid}]
  table[row sep=crcr]{%
-5	0.713846556019543\\
0	1.57385929355615\\
5	3.14221388409291\\
10	5.05430631095684\\
15	7.37455064975869\\
20	9.82734317013133\\
25	11.7561203083318\\
30	13.5044379066454\\
};
\addlegendentry{heuristic, $\sigma_{\xi}^2=8\cdot 10^{-5}$}

\end{axis}

\end{tikzpicture}%
	\caption{Achievable rate of robust and heuristic precoder in case of gaussian distributed position uncertainty}
	\label{fig:robust_gauss}
\end{figure}
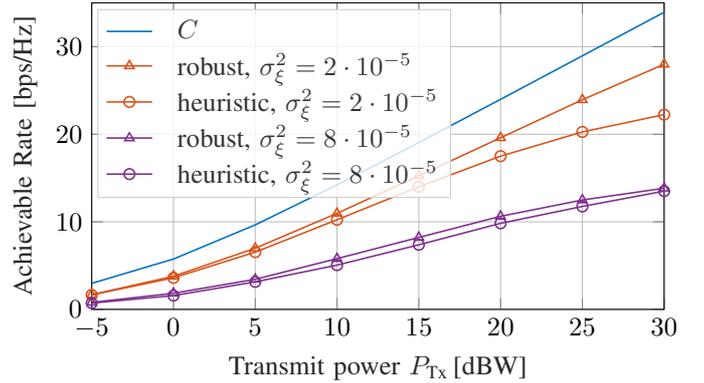

In Fig.~\ref{fig:robust_gauss}, the corresponding performance for Gaussian distribution, i.e., $\erraod_\ell^\xdim,\erraod_\ell^\ydim\sim\mathcal{N}(0,\sigma_{\erraod}^2)$, is shown. The variance $\sigma_{\erraod}^2$ is chosen such that it is almost the same as for the uniformly distributed error in Fig.~\ref{fig:robust_uniform}. It can be seen that the performance gain of the robust precoder compared to the heuristic precoder is smaller than for uniformly distributed position uncertainty. On the one hand, the slope for the robust precoder is not parallel to the channel capacity anymore. Instead, the sum rate degrades stronger for high transmit powers. On the other hand, the impact of gaussian distributed position uncertainty seem to be less severe for the heuristic precoder.


\section{Conclusion}\label{sec:conclusion}
In this paper, a novel robust precoder for \ac{los} communication is presented. Instead of full \ac{csi}, the proposed precoder is based on the second order statistics of the channel, which include imperfect position knowledge of the receivers, as well as statistical knowledge of the position uncertainty and the long term fading of the channel. The position uncertainty of the receivers induce a correlated phase error among the transmit antennas. It has been shown that the resulting statistic of the phase error and the channel statistics are connected via the \ac{cf} of the phase error distribution.
Furthermore, the proposed appraoch to transmit data from a \ac{tx} to multiple satellite has low complexity and is capacity achieving for perfect position knowledge of the satellites and sufficiently large distances between them.

\balance

\section*{Acknowledgment}
This research was supported in part by the German Federal Ministry of Education and Research (BMBF) within the project Open6GHub under grant number 16KISK016A, the German Research Foundation (DFG) under Germany's Excellence Strategy (EXC 2077 at University of Bremen, University Allowance) and by the European Space Agency (ESA) within the SatNEx V activity WI Y2.2-A.

\appendices
\section{Characteristic function}\label{sec:cf}
\paragraph{Uniform Distribution}
Let $\erraod_\text{uni}\sim \mathcal{U}(-\erraoa_{\max}, \erraoa_{\max})$ be uniformly distributed in the interval $[-\erraod_{\max}, \erraod_{\max}]$. Then, the characteristic function $\cfuni$ is a sinc-function, i.e.,
\begin{subequations}
	\begin{align}
		\cfuni(t) &=  \int_{-\erraod_{\max}}^{\erraod_{\max}} \frac{1}{2\erraod_{\max}} e^{j\erraod t} d\erraod \\
		&= \frac{\sin\left(t\erraod_{\max}\right)}{t\erraod_{\max}} = \sinc{t\erraod_{\max}}.
	\end{align} 
\end{subequations}

\paragraph{Gaussian Distribution}
Let $\erraod_{\text{gau}}\sim \mathcal{N}(0, \sigma_{\erraod}^2)$ be gaussian distributed. Then the characteristic function $\cfgau$ follows also a gaussian function \cite[Example~8.5]{book:Kobayashi.Mark.Turin.2011}, i.e.,
	\begin{align}
		\cfgau(t) = \int_{-\infty}^{\infty} \frac{1}{\sigma_{\erraod}\sqrt{2\pi}} e^{-\left(\frac{\erraod}{\sigma_{\erraod}}\right)^2} e^{j\erraod t} d\erraod 
		= e^{-\frac{t^2\sigma^2}{2}}.
	\end{align}


\bibliographystyle{./lib/IEEEtran}
\bibliography{./lib/IEEEabrv,./lib/references_short}

\end{document}